\documentclass[aps,pra,twocolumn]{revtex4-1} 

\usepackage{verbatim}
\usepackage{dsfont} 
\usepackage{amsmath}
\usepackage{amsthm} 
\usepackage{enumerate}
\usepackage{graphicx}
\usepackage{mathtools} 
\usepackage{bm} 
\usepackage{amssymb} 
\usepackage{hyperref} 
\hypersetup{
    colorlinks=true,
    linkcolor=blue,
    citecolor=blue,
    unicode=true,          
}

\def\equationautorefname~#1\null{#1\null}
\usepackage{blindtext}

\newcommand{\ket}[1]{\left| #1 \right\rangle}
\newcommand{\bra}[1]{\left\langle #1 \right|}

\newcommand{\ketbra}[2]{\left|#1 \rangle \langle #2 \right|}
\newcommand{\norm}[1]{\left\lVert#1\right\rVert}
\DeclareMathOperator{\Tr}{Tr}

\DeclareMathOperator*{\argmin}{arg\,min}

\makeatletter
\newcommand{\oset}[3][0ex]{%
  \mathrel{\mathop{#3}\limits^{
    \vbox to#1{\kern-2\ex@
    \hbox{$\scriptstyle#2$}\vss}}}}
\makeatother

\begin{document}



\title{Operational interpretation of weight-based resource quantifiers \\ in convex quantum resource theories of states}


\author{Andr\'es F. Ducuara$^{1,2}$, Paul Skrzypczyk$^{3}$} 

\address{$^{1}$Quantum Engineering Centre for Doctoral Training, \\H. H. Wills Physics Laboratory and Department of Electrical \& Electronic Engineering, University of Bristol, BS8 1FD, UK}

\address{$^{2}$Quantum Engineering Technology Labs, H. H. Wills Physics Laboratory and \\
Department of Electrical \& Electronic Engineering, University of Bristol, BS8 1FD, UK}

\address{$^{3}$H.H. Wills Physics Laboratory, University of Bristol, Tyndall Avenue, Bristol, BS8 1TL, United Kingdom}


\date{\today}

\begin{abstract}
We introduce the resource quantifier of \emph{weight of resource} for convex quantum resource theories of states with arbitrary resources. We show that it captures the advantage that a resourceful state offers over all possible free states, in the operational task of exclusion of subchannels. Furthermore, we introduce information-theoretic quantities related to exclusion and find a connection between the weight of resource of a state, and the exclusion-type information of ensembles it can generate. These results provide support to a recent conjecture made in the context of convex quantum resource theories of measurements, about the existence of a weight-exclusion correspondence whenever there is a robustness-discrimination one. The results found in this article apply to the resource theory of entanglement, in which the weight of resource is known as the best-separable approximation or Lewenstein-Sanpera decomposition, introduced in 1998. Consequently, the results found here provide an operational interpretation to this 21 year-old entanglement quantifier.
\end{abstract} 

\pacs{}

\maketitle

\section{Introduction}

The framework of \emph{Quantum Resource Theories (QRTs)} \cite{RT_review} has proven to be a powerful framework within quantum information. The all-encompassing nature of QRTs has led to the cross fertilisation of ideas amongst different quantum phenomena: Inter-convertibility of states with respect to a particular resource \cite{BG2015}, general resource distillation \cite{BR2019}, impossibility (no-go) theorems \cite{FL2019}, general laws for inter-convertibility of resources \cite{QRT_FL}, insights into general probabilistic theories \cite{RT2}, amongst others \cite{RT_review}. Within the the language of QRTs, properties of different objects deemed as \emph{resources} can be addressed under the same umbrella and therefore, results in a particular QRT with a particular resource has led to insights into different resources and QRTs of different objects.

There are QRTs addressing quantum objects like: states \cite{RT_review}, measurements \cite{RT_measurements0, RT_measurements1, RT_measurements2}, behaviours or boxes \cite{RT_nonlocality, RT_noncontextuality}, steering assemblages \cite{RT_steering}, teleportation assemblages \cite{RoT} and channels \cite{Citeme3, RT_channels1, RT_channels2}. Properties of these objects that are deemed as resources include: entanglement \cite{RoE}, nonlocality \cite{RoNL_RoS_RoI}, steering \cite{RoS}, asymmetry \cite{RoA}, coherence \cite{RoC}, informativeness \cite{RoM}, projective simulability \cite{RT_PS}, incompatibility \cite{RoNL_RoS_RoI, RT_channels1, Citeme1, Citeme2}, teleportation \cite{RoT}, superposition \cite{RT_superposition}, purity \cite{RoP}, magic \cite{RT_magic}, nongaussianity \cite{RT_nongaussianity}, nonmarkovianity \cite{RT_nonmarkovianity1, RT_nonmarkovianity2, Namit}, athermality \cite{RT_thermodynamics}, and reference frames \cite{RT_RF}. 

One of the main goals of QRTs is to define \emph{resource quantifiers} in order to properly quantify the amount of a resource present in an object, as well as to devise \emph{operational tasks} explicitly harnessing these resources. Two prominent families of resource quantifiers are the geometric quantifiers known as: \emph{robustness-based} \cite{RoE,GRoE,RoS,RoA,RoC,RoT,RoT2,RT_magic} and \emph{weight-based} \cite{EPR2,WoE,WoS,WoI_MP, RoNL_RoS_RoI,WoAC} quantifiers. It has been proven that there is a general correspondence between robustness-based resource quantifiers and discrimination-based operational tasks amongst QRTs of various objects \cite{RT1,RT2,RoM}. Recently, it has also been proven \cite{DS2019a}, for the QRT of measurement informativeness, that there exists a parallel quantifier-task correspondence that connects the resource quantifier of weight of informativeness with the operational task of state exclusion, and it was conjectured that this holds true for general resources and convex QRTs of different objects. 

In this work we prove that this conjecture holds true in the context of general convex QRTs of states with arbitrary resources. Specifically, we consider the resource quantifier of \emph{weight of resource} and prove that it quantifies the advantage that a resourceful state offers, when compared to all possible free states, in the operational task of \emph{subchannel exclusion}. In particular, this result holds true when considering the resource of entanglement and therefore, provides an operational interpretation to the weight of entanglement, which is better known as the best separable approximation, or Lewenstein-Sanpera decomposition, introduced in 1998 \cite{WoE}. 

The results presented here nicely complement the \emph{weight-exclusion} correspondence found in \cite{DS2019a} within the QRT of measurement informativeness and therefore, support the conjecture made in the same article about the existence of such a correspondence for convex QRTs of arbitrary objects and arbitrary resources. Interestingly, we will furthermore show that it is  possible to extend the full three-way correspondence found in \cite{DS2019a}, which now links weight and exclusion to the so-called `excludible' information. Explicitly, we prove that for convex QRTs of states with arbitrary resources, the weight of resource also quantifies the single-shot excludible information of an appropriately defined ensemble of states. 

\section{Convex quantum resource theories of states and resource quantifiers}

We start by addressing convex QRTs of states with an arbitrary resource. 

{\it
\textbf{Definition 1:} (Convex QRT of states \cite{RT_review}) Consider the set of quantum states in a Hilbert space of dimension $d$. Consider a property of these states defining a closed convex set which we will call the set of free states and denote as ${\rm F}$. We say a state $\rho \in {\rm F}$ is a \emph{free state}, and $\rho \notin {\rm F}$ is a \emph{resourceful} state.
}

There are numerous properties of quantum states considered as resources namely; entanglement, asymmetry, coherence, amongst many others \cite{RT_review}. We now want to quantify the amount of resource present in an state. We define a weight-based quantifier for an arbitrary resource. 

{\it
\textbf{Definition 2:} (Weight of resource) Consider a convex QRT of states with an arbitrary resource. The weight of resource of a state is given by:
\begin{align}
    {\rm WoR}\left(\rho\right)=
    \min_{\substack{w \geq 0,\\
    \sigma \in {\rm F}, \\
    \rho_G }}
    \left\{ 
     w\,\Big| \, \rho=w\rho_G+(1-w)\sigma
    \right\}.
    \label{eq:WoR}
\end{align}
This quantifies the minimal amount of a general state $\rho_G$ that has to be used to recover the state $\rho$.
}

This quantifier was originally introduced in \cite{EPR2} in the context of nonlocality and independently rediscovered later on in \cite{WoE} within the context of entanglement. It has been addressed under several different names such as: part, content, cost and weight. In this work we use the term \emph{weight} in order to be consistent with recent literature. We now move on to operational tasks.

\section{State exclusion games}

The operational task of state exclusion was first explicitly defined in \cite{CES1}. This task was implicitly introduced in the treatment of the Pusey-Barrett-Rudolph (PBR) theorem \cite{PBR}, in which an ensemble of states allowing conclusive (or perfect) state exclusion was deemed to have the property of Post-Peierls (PP) incompatibility \cite{PP}, or not-PP compatibility \cite{DSR2017}, or antidistinguishability \cite{Leifer}. The task of state exclusion has been explored under noisy channels \cite{DSR2017}, from a communication complexity point of view \cite{CES2, CES3}, for pure states \cite{HK1, HK2}, as well as perfect state exclusion \cite{AM2019}.

{\it
\textbf{Game 1:} (State exclusion \cite{CES1}) A referee has a collection of states $\{\rho_x\}$, $x\in\{1,...,k\}$, and promises to send a player one of these states $\rho_x$ with probability $p(x)$. The goal is for the player to output a guess $g \in \{1,...,k\}$ for a state that was \textit{not} sent. That is, the player succeeds at the game if $g \neq x$ and fails when $g=x$. A given state exclusion game is fully specified by an ensemble $\mathcal{E} = \{\rho_x,p(x)\}$. 
}

The operational task of state exclusion can then be seen as being the opposite to the standard discrimination task, in which the goal is to identify (discriminate), as oppose to exclude. We now consider a quantum protocol for a player to address this task.

{\it
\textbf{Quantum protocol 1:} We consider that the player performs a quantum measurement $\mathbb{M}=\{M_a\}$, $M_a\geq 0$, $\forall a$, $\sum_a M_a=\mathds{1}$ with $o$ outcomes and uses this to simulate a measurement \cite{RoM} $\mathbb{N}=\{N_x\}$ with $k$ outcomes as $N_x=\sum_a q(x|a)M_a$ in order to output the guess of which state to exclude. The probability of error following this strategy is \cite{CES1}:
\begin{align}
    P_{\rm err}^{\rm Q}(\mathcal{E},\mathbb{M}) =
    \min_{\mathbb{N} \preceq \mathbb{M}} 
    \sum_{x} p(x) {\rm Tr}[N_x\rho_x],
    \label{eq:QSE}
\end{align}
with the minimisation being performed over all POVMs $\mathbb{N}$ that are simulable by $\mathbb{M}$ \cite{RoM,DS2019a}.
}

Naturally, we are interested in minimising this probability of error by implementing an optimal POVM. If we consider a binary ensemble, we have that a QSE game is equivalent to a quantum state discrimination (QSD) game and therefore we have $P_{\rm err}^{\rm QSE}(\mathcal{E},\mathbb{M})=P_{\rm err}^{\rm QSD}(\mathcal{E},\mathbb{M})$. Having this, we can then use the Holevo-Helstrom theorem \cite{HHthm1, HHthm2, HHthm3} to address state exclusion games with \emph{binary} ensembles.

{\it 
\textbf{Lemma 1:} (Holevo-Helstrom for state exclusion) The minimum probability of error over all possible POVMs in a state exclusion game with a binary ensemble $\mathcal{E}=\{\rho_x,p(x)\}$ $x\in\{0,1\}$ is given by:
\begin{align}
    \underset{\mathbb{M}}{{\rm min}}\,
    P_{\rm err}^{\rm Q}(\mathcal{E},\mathbb{M})=
    \frac{1}{2}
    \Big(
    1-
    \norm{\tilde \rho_0-\tilde \rho_1}_1
    \Big),
    \label{eq:Lemma1}
\end{align}
with $\tilde \rho_x=p(x)\rho_x$ and the trace norm $\norm{X}_1=\Tr(\sqrt{X^\dagger X})$.
}

The proof of this Lemma is in Appendix A. This result then compares with the standard Holevo-Helstrom theorem for binary QSD \cite{HHthm1, HHthm2, HHthm3} which is usually stated as the maximum probability of succeeding in a binary QSD game being given by $\max_{\mathbb{M}}\,
P_{\rm succ}^{\rm QSD}(\mathcal{E},\mathbb{M})=
\frac{1}{2}
\big(
1+
\norm{\tilde \rho_0-\tilde \rho_1}_1
\big)$. We have that for a binary ensemble, state exclusion is precisely the opposite to state discrimination. This however, does not scale when considering ensembles with more than two states, since $k$-state exclusion games can naturally be defined \cite{CES1}. This Lemma is going to prove useful when addressing one of our main results. We now consider a variant of these state exclusion games.

\section{Subchannel exclusion games}

In analogy to subchannel discrimination games \cite{RoS, RT1}, we now define subchannel exclusion games as follows.

{\it
\textbf{Game 2:} (Subchannel exclusion) The player sends a quantum state $\rho$ to the referee who has a collection of subchannels $\Psi=\{\Psi_x\}$, $x\in\{1,...,k\}$. The subchannels $\Psi_x$ are completely-positive (CP) trace-nonincreasing linear maps such that $\Lambda=\sum_x \Psi_x$ forms a completely-positive trace-preserving (CPTP) linear map. The referee promises to apply one of these subchannels on the state $\rho$ and the transformed state is then sent back to the player. The player then has access to the ensemble $\mathcal{E}_{\Psi}=\{\rho_x, p(x)\}$ with $p(x)=\Tr[\Psi_x(\rho)]$, $\rho_x=\Psi_x(\rho)/p(x)$. The goal is for the player to output a guess $g \in \{1,...,k\}$ for a subchannel that did \textit{not} take place. That is, the player succeeds at the game if $g \neq x$ and fails when $g=x$. 
}

This game can alternatively be seen as playing a quantum state exclusion game with the ensemble $\mathcal{E}^{\rho}_{\Psi}=\{\rho_x, p(x)\}$, in which the player has a certain level of control over the ensemble when proposing the state $\rho$. A particular case of subchannel exclusion is \emph{channel exclusion}, in which $\Psi=\{\Lambda_x,p(x)\}$ with $\{\Lambda_x\}$ being CPTP maps and $p(x)$ a probability distribution. We now consider a quantum protocol for the player to address this game.

{\it
\textbf{Quantum protocol 2:} Consider a subchannel exclusion game in which the player sends a state $\rho$ to the referee who in turn, applies a subchannel from the collection  $\Psi=\{\Psi_x\}$ with $x\in\{1,...,k\}$. Having received the state back, the player now performs a quantum measurement $\mathbb{M}=\{M_a\}$ with $o$ outcomes, and uses this to simulate a measurement $\mathbb{N}=\{N_x\}$ with $k$ outcomes to produce a guess of which subchannel to exclude. The probability of error in quantum subchannel exclusion following this protocol is given by:
\begin{align}
    P_{\rm err}^{\rm Q}(\Psi,\mathbb{M},\rho)=
    \min_{\mathbb{N} \preceq \mathbb{M}} 
    \sum_{x} {\rm Tr}\big[
    N_x\Psi_x(\rho)
    \big],
    \label{eq:QSCE}
\end{align}
with the minimisation being performed over all POVMs $\mathbb{N}$ that are simulable by $\mathbb{M}$ \cite{RoM, DS2019a}. This game can alternatively be seen as playing quantum state exclusion with the ensemble  $\mathcal{E}^{\rho}_{\Psi}=\{\rho_x, p(x)\}$ with $p(x)=\Tr[\Psi_x(\rho)]$, $\rho_x=\Psi_x(\rho)/p(x)$ and we indeed have $P_{\rm err}^{\rm Q}(\Psi,\mathbb{M},\rho)=P_{\rm err}^{\rm Q}(\mathcal{E}^{\rho}_{\Psi},\mathbb{M})$ with the latter as in (\autoref{eq:QSE}).
}

Similarly to state exclusion, we are interested in minimising this probability of error. We will be particularly interested in the performance of a resourceful state compared to the best free state when playing subchannel exclusion games.

\section{All resourceful states are useful in a subchannel exclusion game}

It has already been proven that any resourceful state is useful in a subchannel discrimination game \cite{RT1}. This result addresses a \emph{binary} discrimination game, and since we have already seen that $P_{\rm err}^{\rm QSE}(\mathcal{E},\mathbb{M})=P_{\rm err}^{\rm QSD}(\mathcal{E},\mathbb{M})$ the result then follows. However, since we are now interested in the probability of error, we will write this in the context of state exclusion games as follows.

{\it  
\textbf{Result 1:} For any resourceful state $\rho \notin {\rm F}$, there exists a subchannel exclusion game $\Psi^{\rho}$ for which playing with the state $\rho$ generates fewer errors when compared with any free state as:
\begin{align}
    \min_\mathbb{M}\,
    P^{\rm Q}_{\rm err}(\Psi^{\rho},\mathbb{M},\rho)
    <
    \min_\mathbb{N}
    \min_{\sigma \in {\rm F}}
    P^{\rm Q}_{\rm err}(\Psi^{\rho},\mathbb{N},\sigma).
    \label{eq:result1}
\end{align}
In the right-hand side the error probability is minimised over all possible free states and all measurements.}

The proof of this result follows from the identification that for binary subchannel games we have $P_{\rm err}^{\rm QSE}(\mathcal{E},\mathbb{M})=P_{\rm err}^{\rm QSD}(\mathcal{E},\mathbb{M})$ together with the exclusion version of the Holevo-Helstrom theorem addressed in the previous section. The full proof of this result is in Appendix B. This result shows that every resourceful state is better that any possible free state when playing a tailored subchannel exclusion task, which turns out to always be binary. We now address how to \emph{quantify} the performance of a resourceful state in subchannel exclusion games.

\section{Weight as  the advantage in subchannel exclusion games}

We are now interested in quantifying the performance of a resourceful state in comparison to all free states when playing subchannel exclusion games.

{\it 
\textbf{Result 2:} Consider a subchannel exclusion game in which the player sends a quantum state $\rho$ to the Referee, who in turns applies a subchannel from the ensemble $\Psi=\{\Psi_x\}$ with $x\in\{1,...,k\}$ before sending the state back to the player. The player then implements a measurement $\mathbb{M}=\{M_a\}$ to simulate $\mathbb{N}=\{N_x\}$ and produce the outcome guess $g\in\{1,...,k\}$ representing the choice of a subchannel to be excluded. Then, the quantum-classical ratio of probability of error in subchannel exclusion is lower bounded by a function involving the weight of resource \eqref{eq:WoR}. Furthermore, there exists a subchannel ensemble $\Psi^\rho$ and a measurement $\mathbb{M}^\rho$ for which the lower bound is tight as follows:
\begin{align}
    1-{\rm WoR}(\rho)= 
    \min_{\Psi, \mathbb{M}} 
    \frac{
    P^{\rm Q}_{\rm err}(\Psi,\mathbb{M},\rho)
    }{\displaystyle
    \min_{\sigma\in {\rm F}}
    P^{\rm Q}_{\rm err}(\Psi,\mathbb{M},\sigma)
    }.
    \label{eq:result2}
\end{align}
}
The proof of this result is in Appendix C. We remark that this result holds true for any property of a quantum states that defines a closed convex subset and therefore, it holds in particular for the weight of entanglement, a.~k.~a. the best separable approximation or Lewenstein-Sanpera decomposition \cite{WoE}, and for the weight of asymmetry \cite{WoAC}. 

We note here that in the game the quantum and classical players are required to use the same measurement (\autoref{eq:result2}). In a different setting, we can alternatively ask for the measurements to be chosen independently. We now explore relaxing this measurement constraint.

\section{Quantum-classical ratio with independent measurements}

We now consider a scenario in which  the quantum and classical players implement independent measurements. 

{\it
\textbf{Result 3:} Consider a state $\rho$ and the optimal dual variable $Y^\rho=\sum y_i \ketbra{e_i}{e_i}$ from the dual formulation of the weight of resource (see \eqref{eq:WoR1}). If there exists a set of unitaries $\{U_x\}$ satisfying i) $\sum_x U_x |e_j\rangle \langle e_j |U_x^\dagger=\mathds{1}$, $\forall j$ and ii) $U_i \sigma U_i^\dagger=U_j \sigma U_j^\dagger$, $\forall \sigma\in {\rm F}$, $\forall i,j$, then, the weight of the resource quantifies the advantage of the resourceful state $\rho$ over all free states in subchannel exclusion with independent measurements as:
\begin{align}
    1-{\rm WoR}(\rho)= 
    \min_{\Psi}
    \frac{\displaystyle
    \min_\mathbb{M}\,
    P^{\rm Q}_{\rm err}(\Psi,\mathbb{M},\rho)
    }{\displaystyle
    \min_{\mathbb{N}}
    \min_{\sigma\in {\rm F}}
    P^{\rm Q}_{\rm err}(\Psi,\mathbb{N},\sigma)
    }. 
    \label{eq:result3}
\end{align}
}
The proof of this result is in Appendix D. An example of a resource that satisfies the necessary conditions of Result 3 is coherence \cite{RT1}. This stronger thus holds for this particular resource. We now address single-shot information-theoretic quantities that are also related to the weight of resource.

\section{Single-shot information theory}

We now introduce an exclusion-based quantity closely related to the accessible information of an ensemble, and show that it too relates to the weight of resource. We are interested in the ability of an ensemble $\mathcal{E} = \{\rho_x, p(x)\}$  to be useful for sending exclusion-type information, for example, information of the form `do not cut the blue wire' \cite{DS2019a}. We assume that the ensemble $\mathcal{E}$ is an encoding of a classical random variable $X$, such that $\rho_x$ encodes $x$, the value that should be avoided. A measurement $\mathbb{M} = \{M_g\}$ will be made, in order to produce a new random variable $G$ (the measurement outcome). Using $G$, a prediction for a value $x'\neq x$ can be made, with the optimal choice being $\argmin_g p(g|x)$, i.e. the least likely value of $x$ given the observed $g$, where $p(g|x) = \Tr[M_g \rho_x]$. The total error probability is $P_\mathrm{err}(X|G) = \sum_x p(x) \min_g p(g|x)$ and the associated entropy, which we call the exclusion entropy is $H_{-\infty}(X|G)_{\mathcal{E},\mathbb{M}} = -\log P_\mathrm{err}(X|G)$, which is the order minus-infinity conditional R\'enyi entropy, and where we have explicitly denoted the encoding-decoding dependence upon $\mathcal{E}$ and $\mathbb{M}$ as a final subscript \cite{2015_Informational_nonequilibrium}. Without access to the ensemble, the error probability would be $P_\mathrm{err}(X) = \min_x p(x)$ (i.e. the best choice for $x'$ is just the least likely value of $x$), and the associated exclusion entropy is $H_{-\infty}(X)_{\mathcal{E}} = -\log P_\mathrm{err}(X)$. The reduction in exclusion entropy leads to an mutual exclusion information, $I_{-\infty}(X:G)_{\mathcal{E},\mathbb{M}} = H_{-\infty}(X|G) - H_{-\infty}(X)$. 

We now focus on ensembles which arise by probabilistic ally applying channels onto a fixed quantum state. In particular, we consider the ensemble of channels $\Psi=\{\Lambda_x, p(x)\}$, where each $\Lambda_x$ is a CPTP map. For a fixed state $\rho$, this leads to an ensemble $\mathcal{E}_\rho^\Psi = \{\Lambda_x(\rho),p(x)\}$. We will then compare all of the ensembles that can arise from a state $\rho$, and the mutual exclusion information they lead to, in comparison to the best free state. We find the following result:

{\it
\textbf{Result 4:} The weight of resource of a quantum state quantifies the maximum  increase in mutual exclusion information as:
\begin{multline}
    \max_{\Psi,\mathbb{M}} 
    \left\{
    I_{-\infty}(X \colon G)_{\mathcal{E}_\rho^\Psi,\mathbb{M}}
    -
    \max_{\sigma \in {\rm F}}
    I_{-\infty}(X \colon G)_{\mathcal{E}_\sigma^\Psi,\mathbb{M}}
    \right\}\\
    =-\log \Big[1-{\rm WoR}(\rho)\Big],
    \label{eq:result4}
\end{multline}
with the maximisation over all ensembles of channels and all measurements. 
}

The proof of this result is in Appendix E. This result therefore establishes, for the QRT of states with arbitrary resources, a \emph{three-way correspondence} between weight-based resource quantifiers, exclusion-based tasks, and single-shot information-theoretic quantities in the form of mutual exclusion information. This supports the conjecture made in \cite{DS2019a}, that whenever there is a robustness-discrimination correspondence, there is a weight-exclusion correspondence and furthermore, it extends the full three-way correspondence found in the context of measurement informativeness \cite{DS2019a}. For completeness we also provide here a new result concerning the robustness of resource for arbitrary convex QRTs of states, which was not included in a recent review \cite{RT2}.

{\it
\textbf{Result 5:} The robustness of resource of a quantum state quantifies the maximum increase in mutual accessible information as:
\begin{multline}
    \max_{\Psi,\mathbb{M}} 
    \left\{
    I_{+\infty}(X \colon G)_{\mathcal{E}_\rho^\Psi,\mathbb{M}}
    -
    \max_{\sigma\in {\rm F}}
    I_{+\infty}(X \colon G)_{\mathcal{E}_\sigma^\Psi,\mathbb{M}}
    \right\}\\
    =\log \Big[1+{\rm RoR}(\rho)\Big],
    \label{eq:result5}
\end{multline}
with the maximisation over all ensembles of channels and all measurements. Please see Appendix F for full definitions.
}

The proof of this result is in Appendix F. This results means that the three-way correspondence found in \cite{DS2019a}, for the QRT of measurement informativeness, can indeed be lifted from measurements to states.  

\section{Conclusions}

In this work we have proven, in the context of convex QRTs of states, that \emph{weight-based} resource quantifiers for \emph{arbitrary resources} capture the advantage that a resourceful state has over all free states, in the operational task of \emph{subchannel exclusion}. As a corollary of this result, we have shown that the best separable approximation/Lewenstein-Sanpera decomposition \cite{WoE} quantifies the advantage that an entangled state has over all separable states, in the task of subchannel exclusion. To the best of our knowledge, this is the first operational interpretation that has been given to this entanglement quantifier. Going forward, it would be interesting to derive a version of our result that allows for independent measurements when comparing resourceful and free states, as was done in \cite{RT1} for the robustness of entanglement. 

The results presented here also support the conjecture made in \cite{DS2019a} that, whenever there is an discrimination-based operational task where a robustness-based resource quantifier plays a relevant role, there is an exclusion-based operational task where a weight-based resource quantifier plays a relevant role as well.  It would also be interesting to address this conjecture for other objects, such as steering assemblages or collections of incompatible measurements. All of these considerations are interesting in themselves, but we leave these for future research.

Furthermore, and beyond the weight-exclusion correspondence, we have provided a third connection to single-shot information-theoretic quantities. In particular, we have shown that the weight of resource of a state is also closely related to an exclusion-version of the accessible information of an ensemble of states, which itself originates from a fixed state. 

The results presented in this work nicely fit within the endeavour of linking resource quantifiers to operational tasks in general convex QRTs. One can go even further and consider general probabilistic theories in which the discrimination-robustness correspondence has already been extended \cite{RT2}. We believe that the results presented in this work can be extended to this regime as well, but we leave this for future research. 

\section*{Note added}

During the development of this work we became aware of a complementary work by R. Uola et. al. \cite{Uola}.


\section*{Acknowledgements}

We would like to thank Noah Linden, Patryk Lipka-Bartosik and Tom Purves for insightful discussions. A.F.D acknowledges support from COLCIENCIAS 756-2016. P.S. acknowledges support from a Royal Society URF (UHQT).

\bibliographystyle{apsrev4-1}
\bibliography{bibliography.bib}

\appendix
\section*{APPENDICES}

\section{Proof of Lemma 1}

{\it
\textbf{Lemma 1:} (Holevo-Helstrom for state exclusion) The minimum probability of error in binary state exclusion over all possible POVMs for a given binary ensemble $\mathcal{E}=\{\rho_x,p(x)\}$ $x\in\{0,1\}$ is given by:
\begin{align}
    \underset{\mathbb{M}}{{\rm min}}\,
    P_{\rm err}^{\rm Q}(\mathcal{E},\mathbb{M})=
    \frac{1}{2}
    \Big(
    1-
    \norm{\tilde \rho_0-\tilde \rho_1}_1
    \Big),
    \label{eq:Lemma1A}
\end{align}
with $\tilde \rho_x=p(x)\rho$ and the trace norm $\norm{X}_1=\Tr(\sqrt{X^\dagger X})$ or Schatten $p=1$-norm.
}
\begin{proof}
In a binary state exclusion game we have to exclude from a binary ensemble of states $\mathcal{E}=\{\rho_0, \rho_1, p(0), p(1)\}$ with $p(0)+p(1)=1$ by using a general POVM $\mathbb{M}=\{M_0,M_1\}$, $M_0, M_1\geq 0$, $M_0+M_1=\mathds{1}$. The probability of error is then (\autoref{eq:QSE}):
\begin{align}
    \nonumber 
    P_{\rm err}^{\rm Q}(\mathcal{E},\mathbb{M})=
    {\rm Tr}(M_0\tilde \rho_0)
    +
    {\rm Tr}(M_1\tilde \rho_1),
\end{align}
with $\tilde \rho=p(x)\rho$. We now define a matrix $T$ as $M_0=\frac{\mathds{1}-T}{2}$ and therefore it satisfies $-\mathds{1}\leq T\leq \mathds{1}$. We have:
\begin{align*}
    P_{\rm err}^{\rm Q}(\mathcal{E},\mathbb{M})
    &={\rm Tr}\left[
    \left(\frac{\mathds{1}-T}{2}\right)\tilde \rho_0 
    \right]
    +
    {\rm Tr}\left[
    \left(\frac{\mathds{1}+T}{2}\right)\tilde \rho_1
    \right].
\end{align*}
Reorganising we get:
\begin{align*}
    P_{\rm err}^{\rm Q}(\mathcal{E},\mathbb{M})=
    \frac{1}{2}
    \left(
    1
    +
    {\rm Tr}\Big[
    T(\tilde \rho_1-\tilde \rho_0)
    \Big]
    \right).
\end{align*}
Minimising over POVMs is equivalent to minimising over matrices $T$ and then:
\begin{align*}
    \underset{\mathbb{M}}{{\rm min}}\,
    P_{\rm err}^{\rm Q}(\mathcal{E},\mathbb{M})
    &=\frac{1}{2}
    \left(
    1
    +
    \min_{-\mathds{1}\leq T\leq \mathds{1}} 
    {\rm Tr}\Big[
    T(\tilde \rho_1-\tilde \rho_0)
    \Big]
    \right),\\
    &=\frac{1}{2}
    \left(
    1
    -
    \max_{-\mathds{1}\leq T\leq \mathds{1}} 
    {\rm Tr}\Big[
    T(\tilde \rho_0-\tilde \rho_1)
    \Big]
    \right).
\end{align*}
The last term is the trace norm in disguise and therefore we get the statement in (\autoref{eq:Lemma1A}).
\end{proof}

\section{Proof of Result 1}

{\it  
\textbf{Result 1:} For any resourceful state $\rho \notin {\rm F}$, there exists a subchannel exclusion game $\Psi^{\rho}$ for which playing with the state $\rho$ generates fewer errors when compared with any free state as:
\begin{align}
    \min_\mathbb{M}\,
    P^{\rm Q}_{\rm err}(\Psi^{\rho},\mathbb{M},\rho)
    <
    \min_\mathbb{N}
    \min_{\sigma \in {\rm F}}
    P^{\rm Q}_{\rm err}(\Psi^{\rho},\mathbb{N},\sigma).
    \label{eq:result1A}
\end{align}
In the right-hand side the error probability is minimised over all possible free states and all measurements.}
\begin{proof}
We are going to consider a binary subchannel exclusion game $\Psi=\{\Psi+,\Psi^-\}$. The probability of error in this binary subchannel exclusion problem is given by the exclusion (as opposed to the discrimination) version of the Holevo-Helstrom theorem which we derived in Appendix A (\autoref{eq:Lemma1A}) so we have:
\begin{align}
    \hspace{-0.2cm} \frac{
    \underset{\mathbb{M}}{{\rm min}}\,
    P^{\rm Q}_{\rm err}(\Psi,\mathbb{M},\rho)
    }{
    \underset{\mathbb{N}}{{\rm min}} \hspace{-0.1cm}
    \underset{\sigma \in {\rm F}}{{\min}}
    P^{\rm Q}_{\rm err}(\Psi,\mathbb{N},\sigma)
    }
    =\frac{
    1-
    \norm{(\Psi^{+}-\Psi^{-} )(\rho)}_1
    }{
    1-
    \norm{(\Psi^{+}-\Psi^{-} )(\sigma^*)}_1
    },
    \label{eq:pr1}
\end{align}
with $\sigma^*$ the optimal free state. If we manage to construct subchannels such that:
\begin{align}
    \norm{(\Psi^{+}-\Psi^{-} )(\sigma^*)}_1
    \leq
    \norm{(\Psi^{+}-\Psi^{-} )(\rho)}_1,
\end{align}
then the statement follows. Let us see how this can be done. Given any $\rho \notin {\rm F}$ and by the hyperplane separation theorem (or Hahn-Banach separation theorem) \cite{FAbook2} we have that there exists a bounded self-adjoint operator $W^\rho$ such that: i) ${\rm Tr}(W^\rho\rho) < 0$ and ii) ${\rm Tr}(W^\rho\sigma) \geq 0$, $\forall \sigma \in {\rm F}$. We now construct the operator $X^\rho=\mathds{1}-\frac{W^\rho}{\norm{W^\rho}_{\infty}}$ which is now satisfying the properties i) $0\leq X^\rho$, ii) $0\leq {\rm Tr}(X^\rho\sigma)\leq 1$, $\forall \sigma \in {\rm F}$ and iii) $1<{\rm Tr}(X^\rho\rho)$ and so we have the inequality:
\begin{align}
    0\leq{\rm Tr}(X^\rho\sigma)\leq1<{\rm Tr}(X^\rho\rho),
    \hspace{0.5cm} 
    \forall \sigma \in {\rm F}.
    \label{eq:inequality}
\end{align}
With this operator $X^\rho$ we now construct an appropriate binary subchannel ensemble $\Psi^{\rho}=\left \{\frac{\Lambda^{\rho}_+}{2},\frac{\Lambda^{\rho}_-}{2} \right \}$ as follows:
\begin{align}
    \nonumber 
    \Lambda_{\pm}^\rho(\eta)= 
    &\left(
    \frac{1}{2}
    \pm
    \frac{{\rm Tr}(X^\rho\eta)}{2\norm{X^\rho}_{\infty}}
    \right)
    \ketbra{0}{0}
    +\\+
    &\left(
    \frac{1}{2}
    \mp
    \frac{{\rm Tr}(X^\rho\eta)}{2\norm{X^\rho}_{\infty}}
    \right)
    \ketbra{1}{1}.
\end{align}
We can check that these operators are trace-preserving and therefore the subchannel game is well-defined. Now because of the way that the subchannels have been constructed we obtain:
\begin{align*}
    \norm{(\Lambda_{+}^\rho-\Lambda_{-}^\rho )(\eta)}_1
    =
    \frac{
    2{\rm Tr}(X^\rho\eta)
    }{
    \norm{X^\rho}_{\infty}
    },
    \hspace{0.5cm} \forall \eta.
\end{align*}
We now for any $\sigma \in {\rm F}$ we have:
\begin{align*}
    \norm{(\Lambda_{+}^\rho-\Lambda_{-}^\rho)(\sigma)}_1
    &=
    \frac{
    2{\rm Tr}(X^\rho\sigma)
    }{
    \norm{X^\rho}_{\infty}
    },\\
    & \leq
    \frac{
    2{\rm Tr}(X^\rho\rho)
    }{
    \norm{X^\rho}_{\infty}
    }
    =
    \norm{(\Lambda_{+}^\rho-\Lambda_{-}^\rho)(\rho)}_1.
\end{align*}
The inequality follows from (\autoref{eq:inequality}). We then obtain that the denominator in (\autoref{eq:pr1}) is less than the numerator and so we obtain the claim in (\autoref{eq:result1A}).
\end{proof}

\section{Proof of Result 2}

In order to prove Result 2 we need the following Lemma.

{\it
\textbf{Lemma 2:} The weight of resource of a state $\rho$ (\autoref{eq:WoR}) can be written as the optimisation problem:
\begin{align}
    {\rm WoR}\left(\rho\right)=
    \max_{Y} \hspace{0.2cm}
    &{\rm Tr}[(-Y)\rho]+1, \label{eq:WoR1}\\
    {\rm s. t.} \hspace{0.2cm}
    & Y\geq 0,\label{eq:WoR2}\\
    &{\rm Tr}[Y\sigma]\geq 1, \hspace{0.4cm} \forall \sigma \in {\rm F}.
    \label{eq:WoR3}
\end{align}
This is the dual SDP formulation of the weight of resource.
}

We are now ready to address the proof of Result 2.

{\it 
\textbf{Result 2:} Consider a subchannel exclusion game in which the player sends a quantum state $\rho$ to the Referee, who in turns applies a subchannel from the ensemble $\Psi=\{\Psi_x\}$ with $x\in\{1,...,k\}$ before sending the state back to the player. The player then implements a measurement $\mathbb{M}=\{M_a\}$ to simulate $\mathbb{N}=\{N_x\}$ and produce the outcome guess $g\in\{1,...,k\}$ representing the choice of a subchannel to be excluded. Then, the quantum-classical ratio of probability of error in subchannel exclusion is lower bounded by a function involving the weight of resource \eqref{eq:WoR}. Furthermore, there exists a subchannel ensemble $\Psi^\rho$ and a measurement $\mathbb{M}^\rho$ for which the lower bound is tight as follows:
\begin{align}
    1-{\rm WoR}(\rho)= 
    \min_{\Psi, \mathbb{M}} 
    \frac{
    P^{\rm Q}_{\rm err}(\Psi,\mathbb{M},\rho)
    }{\displaystyle
    \min_{\sigma\in {\rm F}}
    P^{\rm Q}_{\rm err}(\Psi,\mathbb{M},\sigma)
    }.
    \label{eq:result2A}
\end{align}
}
\begin{proof}
    Let us start by proving that for a given $\rho$, $1-{\rm WoR}(\rho)$ places a lower bound to the quantum-classical ratio in any subchannel exclusion game, that this, for any tuple $(\mathbb{M}, \Psi)$. Given $\rho$ and any tuple $(\Psi, \mathbb{M})$ we have:
    \begin{align}
        \nonumber &P^{\rm Q}_{\rm err}(\Psi,\mathbb{M},\rho)=\\
        \nonumber &=\min_{\mathbb{N} \preceq \mathbb{M}}
        \sum_{x} {\rm Tr}[N_x\Psi_x(\rho)],\\
        \nonumber &\geq \Big[1-{\rm WoR}(\rho)\Big] 
        \min_{\mathbb{N} \preceq \mathbb{M}}
        \sum_{x} {\rm Tr}[N_x\Psi_x(\sigma^*)],\\
        \nonumber &\geq \Big[1-{\rm WoR}(\rho)\Big]
        \min_{\sigma \in {\rm F}}
        \min_{\mathbb{N} \preceq \mathbb{M}}
        \sum_{x} {\rm Tr}[N_x\Psi_x(\sigma)],\\
        & = \Big[1-{\rm WoR}(\rho)\Big]
        \min_{\sigma \in {\rm F}}
        P^{\rm Q}_{\rm err}(\Psi,\mathbb{M},\sigma).
        \label{eq:result1a}
    \end{align}
    In the first inequality we used (\autoref{eq:WoR}) from which we get $\rho \geq [1-{\rm WoR}(\rho)]\sigma^*$ and since $\Psi_x$ are linear maps we have $\Psi_x(\rho)\geq [1-{\rm WoR}(\rho)] \Psi_x(\sigma^*)$, $\forall x$. In the second inequality we allow ourselves to minimise over all free states. We now address how to achieve the lower bound. Consider a given $\rho$ and let us construct an appropriate subchannel exclusion game $\Psi^\rho=\{\Psi_x^\rho\}$ and an appropriate measurement $\mathbb{M}^\rho=\{M_x^\rho\}$ achieving the lower bound.  Let us start by considering the optimal matrix coming out of the Dual SDP formulation of the weight of resource (\autoref{eq:WoR1}) as $Y^\rho$. By spectral decomposition we have $Y^\rho=\sum_{i=1}^d y_i \ketbra{e_i}{e_i}$ with $y_i \in \mathds{R}$ and $\{\ket{e_i}\}$ an orthonormal basis. We now consider a set of unitary matrices $\{U_x\}$, $x\in\{1,...,d\}$ satisfying the property $\sum_x U_x e_j U_x^\dagger=\mathds{1}$, $\forall j$, $e_j=\ketbra{e_j}{e_j}$. This can be done by defining, for instance, $U_y=\sum_{x=1}^d \ket{e_{x+y}}\bra{e_x}$. We now define a subchannel game $\Psi^\rho=\{\Psi^\rho_x\}$ and a measurement $\mathbb{M}^\rho=\{M_x^\rho\}$ as:
    \begin{align}
        \Psi^\rho_x(.)&=\frac{1}{d}U_x(.)U_x^\dagger,
        \label{eq:SC2}\\
        M^\rho_x&=\frac{1}{\Tr (Y^\rho)}U_x Y^\rho U_x^\dagger.
        \label{eq:M2}
    \end{align}
One can check that the subchannels and the measurement are well defined. We can now check that the probability of error in subchannel exclusion for a state $\eta$ is given by:
    \begin{align*}
        P^{\rm Q}_{\rm err}(\Psi^\rho,\mathbb{M}^\rho,\eta)=
        \frac{
        \Tr(Y^\rho\eta)
        }{
        \Tr(Y^\rho)
        }, \hspace{0.5cm}
        \forall \eta.
        \end{align*}
The quantum-classical ratio then satisfies:
    \begin{align}
        \nonumber \frac{
        P^{\rm Q}_{\rm err}(\Psi^\rho,\mathbb{M}^\rho,\rho)
        }{
        \min_{\sigma \in {\rm F}}
        P^{\rm Q}_{\rm err}(\Psi^\rho,\mathbb{M}^\rho,\sigma)
        }
        &=
        \frac{\Tr(Y^\rho\rho)
        }{
        \nonumber \Tr(Y^\rho \sigma^*)},\\
        &\leq
        \Tr(Y^\rho\rho),\\
        &=1-{\rm WoR}(\rho).
        \label{eq:previous}
    \end{align}
The inequality follows because of (\autoref{eq:WoR3}) and the last equality because of (\autoref{eq:WoR1}). The inequality \autoref{eq:previous} together with (\autoref{eq:result1a}) leads to:
\begin{align*}
        \frac{
        P^{\rm Q}_{\rm err}(\Psi^\rho,\mathbb{M}^\rho,\rho)
        }{
        \min_{\sigma \in {\rm F}}
        P^{\rm Q}_{\rm err}(\Psi^\rho,\mathbb{M}^\rho,\sigma)
        }
        =1-{\rm WoR}(\rho).
\end{align*}
Therefore, for any given $\rho$ we can find both a subchannel exclusion game $\Psi^\rho=\{\Psi^\rho_x\}$ and a measurement $\mathbb{M}^\rho=\{M^\rho_x\}$, given by (\autoref{eq:SC2}) and (\autoref{eq:M2}) respectively, such that the the quantum-classical ratio achieves the lower bound and therefore obtaining the claim in (\autoref{eq:result2A}).
\end{proof}

\section{Proof of Result 3}

In order to prove Result 3 we need the following theorem, which is Theorem 1 in \cite{CES1}.

{\it
\textbf{Theorem:} (Necessary and sufficient conditions for optimality in state exclusion \cite{CES1}) Consider a state exclusion game defined by the ensemble $\mathcal{E}=\{\rho_x, p(x)\}$. We now consider a measurement $\mathbb{M}=\{M_x\}$, and the operator:
\begin{align}
    N=\sum_x \tilde \rho_x M_x.
\end{align}
The measurement $\mathbb{M}=\{M_x\}$ is an optimal measurement for playing quantum state exclusion with the ensemble $\mathcal{E}=\{\rho_x,p(x)\}$ if and only if the operator $N$ satisfies the following two conditions:
\begin{align}
    &\text{i)}\, N^\dagger=N,\label{eq:oc1}\\
    &\text{ii)}\, \tilde \rho_x-N \geq 0,  \hspace{0.5cm} \forall x, \label{eq:oc2} 
\end{align}
with $\tilde \rho=p(x)\rho_x$.
}

We now are ready to address Result 3.

{\it
\textbf{Result 3:} Consider a state $\rho$ and its associated matrix from (\autoref{eq:WoR1}) written in spectral decomposition as $Y^\rho=\sum y_i \ketbra{e_i}{e_i}$. If there exist a set of unitaries $\{U_x\}$ satisfying the two conditions:
\begin{align}
    &\sum_x U_x e_jU_x^\dagger=\mathds{1}, \hspace{0.5cm} \forall j, \label{eq:c1}\\
    &U_i \sigma U_i^\dagger=U_j \sigma U_j^\dagger, \hspace{0.3cm} \forall \sigma\in {\rm F},\hspace{0.3cm} \forall i,j.
    \label{eq:c2}
\end{align}
Then, the weight of the resource quantifies the advantage of the resourceful state $\rho$ over all free states in quantum subchannel exclusion games with independent measurements as:
\begin{align}
    1-{\rm WoR}(\rho)= 
    \min_{\Psi}
    \frac{
    \underset{\mathbb{M}}{\min}\,
    P^{\rm Q}_{\rm err}(\Psi,\mathbb{M},\rho)
    }{
    \underset{\mathbb{N}}{\min}\,
    \underset{\sigma\in {\rm F}}{{\rm min}}\,
    P^{\rm Q}_{\rm err}(\Psi,\mathbb{N},\sigma)
    }, 
    \label{eq:result3A}
\end{align}
with the probability of error in subchannel exclusion (\autoref{eq:QSCE}) and the weight of resource (\autoref{eq:WoR}).
}

The proof of this result uses similar techniques to the discrimination case in \cite{RT2}. The subtlety lies in that we now need to check the necessary and sufficient conditions for quantum state exclusion \cite{CES1}, as opposed to those for quantum state discrimination \cite{RT2}. We explicitly write down the proof for completeness.

\begin{proof}
    The lower bound can be proven similarly as in Result 2, so we only address here how to achieve the lower bound. Similarly to Result 2, we need to define an optimal subchannel game and a measurement. We are going to define them similarly to Result 2 and we will see that this measurement turns out to be optimal when considering separable states, meaning that conditions (\autoref{eq:oc1}) and (\autoref{eq:oc2}) are satisfied. We now define the subchannels and measurement as:
    \begin{align}
        \Psi^\rho_x(.)&=\frac{1}{d}U_x(.)U_x^\dagger,\label{eq:SC3}\\
        M^\rho_x&=\frac{1}{\Tr (Y^\rho)}U_x Y^\rho U_x^\dagger,
        \label{eq:M3}
    \end{align}
    with $\{U_x\}$ as described in the statement of Result 3. Checking that they satisfy optimality condition 1 (\autoref{eq:oc1}) when considering free states $\sigma$. We now look at the subchannel exclusion game as a state exclusion game with $\mathcal{E}^{\sigma^*}_{\Psi^\rho}=\left \{\frac{1}{d}, \sigma^x\right \}$ with $\sigma^x=U_x \sigma^* U_x^\dagger$, $\sigma^*$ being the optimal free state. We note that the assumption (\autoref{eq:c2}) translates now to $\sigma^i=\sigma^j$ $\forall i,j$. We now want to argue that the measurement in (\autoref{eq:M3}) is optimal for this state exclusion game. We now calculate the operator $N$ and check the first optimality condition (\autoref{eq:oc1}). We have that:
    \begin{align*}
        M_i\left(
        \frac{1}{d}\sigma^i-\frac{1}{d}\sigma^j
        \right)
        M_j=0 
        \hspace{0.5cm}
        \forall i,j.
    \end{align*}
    because the quantity inside the parenthesis is always zero. These conditions imply that $N^\dagger=N$ as desired, let us see this. The previous is equivalent to:
\begin{align*}
        \frac{1}{d} M_i\sigma^iM_j
        =
        \frac{1}{d} M_i\sigma^jM_j, \hspace{0.5cm} \forall i,j.
\end{align*}
Adding over $j$ we have:
\begin{align*}
        \frac{1}{d} M_i\sigma^i
        =
        \frac{1}{d} M_i\sum_j\sigma^jM_j.
\end{align*}
Adding now over $i$ we have:
\begin{align*}
        N^\dagger=\sum_i M_i\sigma^i \frac{1}{d}
        =
        \sum_j \frac{1}{d} \sigma^jM_j=N.
\end{align*}
Therefore we have that the optimality condition (\autoref{eq:c1}) is satisfied. We take into account that $M-i$ and $\rho$ are positive operators so the are self-adjoint. We now check the second optimality condition (\autoref{eq:oc2}). We have:
    \begin{align*}
       \tilde \rho_x-N&=
       \frac{1}{d} \sigma^x-\sum_y \frac{1}{d}\, \sigma^y \, M_y, \\
       &=\frac{1}{d} \sigma^x \left(
       \mathds{1}-\sum_y M_y
       \right)=0\geq 0, \hspace{0.3cm} \forall x.
    \end{align*}
    In the second line we have used that $\sigma^x=\sigma^y$, $\forall x,y$ which is the assumption (\autoref{eq:c2}). Therefore the measurement (\autoref{eq:M3}) is an optimal measurement for quantum state exclusion and we obtain the statement in (\autoref{eq:result3A}).
\end{proof}

\section{Proof of Result 4}

{\it
\textbf{Result 4:} The weight of resource of a quantum state quantifies the maximum  increase in mutual exclusion information as:
\begin{multline}
    \max_{\Psi,\mathbb{M}} 
    \left\{
    I_{-\infty}(X \colon G)_{\mathcal{E}_\rho^\Psi,\mathbb{M}}
    -
    \max_{\sigma \in {\rm F}}
    I_{-\infty}(X \colon G)_{\mathcal{E}_\sigma^\Psi,\mathbb{M}}
    \right\}\\
    =-\log \Big[1-{\rm WoR}(\rho)\Big],
    \label{eq:result4A}
\end{multline}
with the maximisation over all ensembles of channels and all measurements. 
}
\begin{proof}
The minus-infinity mutual information between classical random variables $X_\Psi$ and $G_\mathbb{M}$ is given by:
\begin{align*}
    I_{-\infty}^\rho(X _\Psi \colon G_\mathbb{M})= H_{-\infty}^\rho(X_\Psi|G_\mathbb{M})-H_{-\infty}(X_\Psi),
\end{align*}
with $H_{-\infty}(X_\Psi)=- \log [\min_x p(x)]$, $H_{-\infty}^\rho(X_\Psi|G_\mathbb{M})=-\log \sum_g \min_x p(g,x)$, $p(g,x)=p(g|x)p(x)$. Using $p(g|x)=\Tr[M_g\Lambda_x(\rho)]$ then $H_{-\infty}^\rho(X_\Psi|G_\mathbb{M})=-\log \sum_g \min_x \Tr[M_g\Psi_x(\rho)]$. Considering $f_g(x)={\rm Tr}[M_g \Psi_x(\rho)]$ and using:
\begin{align*}
    \min_x f_g(x) =\min_{\{p(x|g)\}} \sum_x p(x|g) f_g(x),
\end{align*}
we have:
\begin{align} 
    \nonumber & H_{-\infty}^\rho(X_\Psi|G_\mathbb{M})\\
    &=- \log 
    \sum_g \min_{\{p(x|g)\}} \sum_x p(x|g) 
    f_g(x)
    ,\\
    \nonumber &=- \log 
    \sum_g \min_{\{p(x|g)\}} \sum_x p(x|g) {\rm Tr}[M_g \Psi_x(\rho)],\\
    \nonumber &=-\log \min_{\{p(x|g)\}} \sum_x {\rm Tr}\left[\left( \sum_g p(x|g) M_g \right) \Psi_x(\rho) \right],\\
    \nonumber &= -\log  
    \min_{\mathbb{N}\prec\mathbb{M}} \sum_x {\rm Tr}[N_x \Psi_x(\rho)]
    ,\\
    &= -\log P^{\rm Q}_{\rm err}(\Psi,\mathbb{M},\rho).
    \label{eq:B3W}
\end{align}
where we have the minimisation over all measurements $\mathbb{N}$ being simulable by $\mathbb{M}$ \cite{simulability, DS2019a}. We now have:
\begin{align*} 
    &I_{-\infty}^\rho(X_\Psi|G_\mathbb{M})
    -
    \max_{\sigma\in {\rm F}}
    I_{-\infty}^\sigma(X_\Psi|G_\mathbb{M})=\\
    &=H_{-\infty}^\rho(X_\Psi|G_\mathbb{M})-
    \max_{\sigma\in {\rm F}(\mathds{H})}
    H_{-\infty}^\sigma(X_\Psi|G_\mathbb{M}),\\
    &=-\log 
    \Big [
    P^{\rm Q}_{\rm err}(\mathcal{E}^\rho_{\Psi},\mathbb{M})
    \Big ]
    - \max_{\sigma\in {\rm F}}
    -\log \Big [
    P^{\rm Q}_{\rm err}(\mathcal{E}^{\sigma}_{\Psi},\mathbb{M})
    \Big ],\\
    &
    =-\log 
    \Big [
    P^{\rm Q}_{\rm err}(\mathcal{E}^\rho_{\Psi},\mathbb{M})
    \Big ]
    + \min_{\sigma\in {\rm F}}
    \log \Big [
    P^{\rm Q}_{\rm err}(\mathcal{E}^{\sigma}_{\Psi},\mathbb{M})
    \Big ],\\
    &=-\left \{ \log 
    \Big [
    P^{\rm Q}_{\rm err}(\mathcal{E}^\rho_{\Psi},\mathbb{M})
    \Big ]
    -
    \min_{\sigma\in {\rm F}}
    \log \Big [
    P^{\rm Q}_{\rm err}(\mathcal{E}^{\sigma}_{\Psi},\mathbb{M})
    \Big ]
    \right \},\\
    &=-\log \left \{
    \frac{
    P^{\rm Q}_{\rm err}(\Psi,\mathbb{M},\rho)
    }{
    \min_{\sigma\in {\rm F}}P^{\rm Q}_{\rm err}(\Psi,\mathbb{M},\sigma),
    }
    \right \}.
\end{align*}
We now maximise over all games $\Psi$ and all measurements $\mathbb{M}$ and get:
\begin{align*} 
    &\max_{\Psi,\mathbb{M}} 
    \left\{
    I_{-\infty}^\rho(X_\Psi|G_\mathbb{M})
    -
    \max_{\sigma\in {\rm F}}
    I_{-\infty}^\sigma(X_\Psi|G_\mathbb{M})
    \right \},\\
    &=\max_{\Psi,\mathbb{M}}
    -\log \left \{
    \frac{
    P^{\rm Q}_{\rm err}(\Psi,\mathbb{M},\rho)
    }{
    \min_{\sigma\in {\rm F}}P^{\rm Q}_{\rm err}(\Psi,\mathbb{M},\sigma),
    }
    \right \},\\
    &=-\min_{\Psi,\mathbb{M}}
    \log \left \{
    \frac{
    P^{\rm Q}_{\rm err}(\Psi,\mathbb{M},\rho)
    }{
    \min_{\sigma\in {\rm F}}P^{\rm Q}_{\rm err}(\Psi,\mathbb{M},\sigma),
    }
    \right \},\\
    &=-
    \log 
    \Big [1-{\rm WoR}(\rho) \Big].
\end{align*}
In the last line we used Result 2.
\end{proof}

\section{Proof of Result 5}

To prove Result 5 we need to introduce the resource quantifier of robustness of a resource as follows.

{\it
\textbf{Definition:} (Generalised robustness of resource \cite{RT1}) Consider a convex QRT of states with an arbitrary resource. The robustness of resource of a state is given by:
\begin{align}
    {\rm RoR}\left(\rho\right)=
    {\scriptsize
    \begin{matrix}
    \text{\small \rm min}\\
    r \geq 0\\
    \sigma \in {\rm F} \\
    \rho_G \\
    \end{matrix}
    }
    \left\{ 
    \rule{0cm}{0.6cm} w\,\bigg| \, \rho+r\rho_G=(1+r)\sigma
    \right\}.
    \label{eq:RoR}
\end{align}
This quantifies the minimum amount of a general state $\rho_G$ that has to be added to $\rho$ such that we get a free state $\sigma$.
}

We also need the following theorem.

{\it 
\textbf{Theorem:} (\cite{RT1}) Consider a subchannel discrimination game in which the Player sends a state $\rho$ to the Referee who in turn applies a subchannel from $\Psi=\{\Psi_x\}$ with $x\in\{1,...,k\}$, before sending it back to the Player. The Player is being asked to produce an outcome guess index $g\in\{1,...,k\}$ representing the choice of a subchannel to be identified. The Player implements a measurement $\mathbb{M}=\{M_a\}$ and simulates the measurement $\mathbb{N}=\{N_x\}$ with $k$ outcomes \cite{RoM, DS2019a}. The quantum-classical ratio of probability of success in state discrimination is upper bounded by a function involving the robustness of resource (\autoref{eq:RoR}) and furthermore, that for fixed state $\rho$, there exists an ensemble of subchannels $\Psi^\rho$ and a measurement $\mathbb{M}^\rho$ in which the upper bound is tight as follows:
\begin{align}
    \underset{\Psi, \mathbb{M}}{{\rm max}}
    \frac{
    P^{\rm Q}_{\rm succ}(\Psi,\mathbb{M},\rho)
    }{
    \underset{\sigma\in {\rm F}}{{\rm max}}\,
    P^{\rm Q}_{\rm succ}(\Psi,\mathbb{M},\sigma)
    } 
    =
    1+{\rm RoR}(\rho),
    \label{eq:thmRT}
\end{align}
with {\rm RoR} the robustness of resource (\autoref{eq:RoR}), and the probability of success in quantum subchannel discrimination \cite{RT1}.
}

The proof of this theorem can be found in \cite{RT1}. We are now ready to prove Result 5. The proof of this result is similar to that of Result 4, but we reproduce it here for completeness. 

{\it
\textbf{Result 5:} The robustness of resource of a quantum state quantifies the maximum increase in mutual accessible information as:
\begin{multline}
    \max_{\Psi,\mathbb{M}} 
    \left\{
    I_{+\infty}(X \colon G)_{\mathcal{E}_\rho^\Psi,\mathbb{M}}
    -
    \max_{\sigma\in {\rm F}}
    I_{+\infty}(X \colon G)_{\mathcal{E}_\sigma^\Psi,\mathbb{M}}
    \right\}\\
    =\log \Big[1+{\rm RoR}(\rho)\Big],
    \label{eq:result5A}
\end{multline}
with the maximisation over all ensembles of channels and all measurements.
}
\begin{proof}
The plus-infinity mutual information between classical random variables $X_\Psi$ and $G_\mathbb{M}$ is given by:
\begin{align*}
    I_{+\infty}^\rho(X_\Psi \colon G_\mathbb{M})=
    H_{+\infty}(X_\Psi )-
    H_{+\infty}^\rho(X_\Psi |G_\mathbb{M}),
\end{align*}
with $H_{+\infty}(X_\Psi)=- \log [\max_x p(x)]$, $H_{+\infty}^\rho(X_\Psi|G_\mathbb{M})=- \log \sum_g \max_x p(g,x)$ with $p(g,x)=p(g|x)p(x)$. We have $p(g|x)=\Tr[M_g \Lambda_x(\rho)]$ and then $H_{+\infty}^\rho(X_\Psi|G_\mathbb{M})=- \log \sum_g \max_x \Tr[M_g\Psi_x(\rho)]$. Considering $f_g(x)={\rm Tr}[M_g \Psi_x(\rho)]$ and using:
\begin{align*}
    \max_x f_g(x) =\max_{\{p(x|g)\}} \sum_x p(x|g) f_g(x),
\end{align*}
we have:
\begin{align} 
    \nonumber & H_{+\infty}(X_\Psi|G_\mathbb{M})\\
    &=- \log 
    \nonumber \sum_g \max_{\{p(x|g)\}} \sum_x p(x|g) 
    f_g(x)
    ,\\
    \nonumber &=- \log 
    \sum_g \max_{\{p(x|g)\}} \sum_x p(x|g) {\rm Tr}[M_g \Psi_x(\rho)],\\
    \nonumber &=-\log \max_{\{p(x|g)\}} \sum_x {\rm Tr}\left[\left( \sum_g p(x|g) M_g \right) \Psi_x(\rho) \right],\\
    \nonumber &= -\log  
    \max_{\mathbb{N}\prec \mathbb{M}} \sum_x {\rm Tr}[N_x \Psi_x(\rho)]
    ,\\
    &= -\log P^{\rm Q}_{\rm succ}(\Psi,\mathbb{M},\rho),
    \label{eq:B3R}
\end{align}
where we have the maximisation over all measurements $\mathbb{N}$ being simulable by $\mathbb{M}$ \cite{simulability, DS2019a}. Considering now the quantity of interest we get:
\begin{align*}
    &I_{+\infty}^\rho(X_\Psi \colon G_\mathbb{M})-
    \max_{\sigma\in {\rm F}}
    I_{+\infty}^\sigma(X_\Psi \colon G_\mathbb{M})\\
    &=-H_{+\infty}^\rho(X_\Psi|G_\mathbb{M})-
    \max_{\sigma\in {\rm F}}
    -H_{+\infty}^\sigma(X_\Psi|G_\mathbb{M}),\\
    &=\log 
    \Big [
    P^{\rm Q}_{\rm succ}(\Psi,\mathbb{M},\rho)
    \Big ]
    - \max_{\sigma\in {\rm F}}
    \log \Big [
    P^{\rm Q}_{\rm succ}(\Psi,\mathbb{M},\sigma)
    \Big ],\\
    &= \log \left \{
    \frac{P^{\rm Q}_{\rm succ}(\Psi,\mathbb{M},\rho)
    }{
    \max_{\sigma\in {\rm F}}
    P^{\rm Q}_{\rm succ}(\Psi,\mathbb{M}, \sigma)
    }
    \right \}.
\end{align*}
Now entering with the maximisation over game settings, measurements and using (\autoref{eq:thmRT}) we obtain the claim in (\autoref{eq:result5A}).
\end{proof}

\end{document}